\documentclass[preprint]{aastex}

\shorttitle{Giant pulse emission from PSR B0950+08}
\shortauthors{Singal and Vats}

\begin{document}

\title{Giant pulse emission from PSR B0950+08}
\author{Ashok K. Singal and Hari Om Vats}
\affil{Astronomy and Astrophysics Division, Physical Research Laboratory,\\
Navrangpura, Ahmedabad - 380 009, India}
\email{asingal@prl.res.in} 
\begin{abstract}
We present here the detection of giant-pulse emission from PSR B0950+08, 
a normal-period pulsar. The observations, made at 103 MHz and lasting for about
ten months, have shown on a number of days the frequency of occurrence of giant pulses to be the highest 
among the known pulsars. The flux--density level of successive giant pulses fluctuates rapidly 
and their occurrence rates within a day's observations as well as between neighboring days 
show large variations. While on some days PSR B0950+08 shows a large number of giant pulses, there are 
other days when it shows only ``quasi--nulls'' with no detectable emission in the power spectrum 
or in the folded pulse data. The cumulative intensity 
distribution of these giant pulses appears to follow a power law, with 
index $-2.2$. After eliminating instrumental, ionospheric, interplanetary 
and interstellar diffractive and refractive scintillation effects as the cause, 
it appears that these intensity variations are intrinsic to the pulsar. 
We suggest that the giant pulse emission and nulling may be opposite manifestations of the same 
physical process, in the former case an enhanced number of charges partaking in the 
coherent radiation process giving rise to an extremely high intensity while in the latter case 
the coherence could be minimal.
\end{abstract}


\section{Introduction}
At radio wavelengths, pulsars show pulse to pulse intensity variations 
which lie typically  within an order
of magnitude of the average pulse strength. However, in about half a dozen pulsars,
the pulse intensity is seen to sometimes exceed the mean pulse strength by much more 
than that. These are called giant pulses. 
With the exception of the Crab pulsar (Lundgren et al. 1995),  
PSR B1937+21 (Cognard et al. 1996) and a handful of other cases (see Knight 2006),  
pulses much stronger than 10 times the mean have not generally been seen, and pulsars with 
giant pulses seem to be rather uncommon (Johnston \& Romani 2002).
Triggered by a chance observation of a large number of giant pulses
on August 8, 1997 (Vats et al. 1997; Singal, Vats \& Deshpande 2000) from the extremely variable
pulsar PSR B0950+08 (Pilkington et al. 1968; Cole, Hess \& Page 1970), we
undertook a long-term monitoring program of this object at 103 MHz by 
observing it daily for half an hour. The aim was to confirm and study the nature of these giant 
pulses from their cumulative intensity distribution and from the fluctuations in their occurrence 
rate as well as long-term  variations, if any, in their daily mean. 
Such statistical knowledge is still lacking, not merely  because only a
few pulsars are known to exhibit giant pulse phenomenon but also because to
monitor such sources on a regular basis to get their temporal statistics it may require a fairly large amount of
precious telescope time, spread over many months or longer, which may be difficult to get allotted.
We found some curious results for PSR B0950+08 in its giant pulse emission.

In recent literature (Johnston et al. 2001; Cairns 2004) a distinction is made between giant pulses and 
giant micropulses, the latter being short--duration intense events that may have 
phase--resolved flux densities more than 10 times the average flux density but 
have pulse--integrated flux density less than 10 times the average, and are thus not called 
giant pulses. The Vela pulsar (Johnston et al. 2001), B1706-44 (Johnston \& Romani 2002), 
J0437-4715 (Jenet et al. 1998) and even B0950+08 (Cairns, Johnston \& Das 2004) have been 
classified as giant micropulse cases. But here we show that B0950+08 emits giant pulses as their 
pulse-integrated flux does exceed the average pulse strength by 10 times or even much more.
\section{Observations}
Our observations were made with the Rajkot radio telescope, situated at
a location (longitude 70.7$^\circ$ E, latitude 22.3$^\circ$ N)
in the western part of India (Vats et al. 1999).   
The telescope is a transit instrument, consisting of 1024 dipoles spread
over a 5000 square meter area, with a maximum antenna spacing of $64 \:\lambda$ in the north-south 
and $7.5 \:\lambda$ in the east-west. The telescope, designed in the 
1970s primarily for the interplanetary scintillation (IPS) studies, operates
at a center frequency of 103 MHz, with a bandwidth of 1.6 MHz, and has only a single polarization.
The wide east-west beam ($\sim 8^\circ$) of the telescope
allows an object near the meridian to be studied for about half an hour,
the north-south pointing is done using the phased array technique. The
sampling time interval selected for IPS observations is 48 ms, with a
receiver time constant of 100 ms. Our initial observations (on August 8, 1997) came from 
regular IPS monitoring of 3C237 which follows PSR B0950+08 in right ascension
by about 15 minutes.
On these records we noted that on some days there were a large number of very 
intense pulses from PSR B0950+08. We decided to monitor
this pulsar more regularly, throughout the year. Though an occurrence of
some very intense pulses from PSR B0950+08, observed at 103 MHz, were reported on
an earlier occasion (Deshpande et al. 1994), the nature of these as
regular giant pulses was not recognized at that time.
For the sake of consistency with our initial observations, 
we decided to continue with the same observational setup.
Since the dispersion measure for PSR B0950+08 is small
(DM $\sim 3$ pc cm$^{-3}$) the expected smearing ($\sim 36$ ms) 
across the band is not too detrimental for the observational setup,
more so as our aim was to study the total pulse intensity.
We obtained a total of 141 days of successful observations during the
period July 1997-May 1998. Useful observations could not be made on other days
due to instrumental problems, presence of heavy radio interference or due 
to a simple shortage of manpower. In these
141 days, with daily 32 minute observing periods, data for more than
a million pulses ($P=253$ ms) have thus been recorded, yielding a mean
pulse intensity of $\sim 3$ Jy. While most individual pulses
were too weak to be detected (signal to noise $\stackrel{<}{_{\sim}}0.2$),
single giant pulses may greatly exceed the noise level.
We have chosen 10 times the mean ($\sim 30$ Jy) as the threshold pulse intensity for
a pulse to be called giant. This definition is consistent with the
usage in the recent literature (Karuppusamy, Stappers, \& van Straten 2010) and safeguards 
against the inclusion of any significant number of spurious giant pulses.
To minimize the effect of radio interference, individual giant pulses were 
identified by their phase;  further we 
rejected all records with a certain amount of discernible radio interference.
Daily calibrations were done before and/or after the observations using 3C196 and 3C273.

\section{Results}
Some details of the giant--pulses statistics are available in the literature, 
but for only a couple of pulsars. 
For the Crab pulsar about 2.5 percent of all pulses were observed to have  
pulse intensity more than 20 times the average value (Lundgren et al.
1995), while Cognard et al. (1996) found that in the millisecond pulsar PSR B1937+21 
about one pulse per ten thousand exceeded 20 times the mean ``pulse--on'' flux density.
Among our approximately one million pulses, the corresponding number  
(i.e., with pulse strength 20 times the mean pulse--on flux density) was found to be $\sim1\%$ of 
the total, while about one in ten thousand was  as large as 100 times  the mean pulse level, 
with some individual pulses exceeding the mean by a factor of 300. 
This makes PSR B0950+08 one of the most active pulsars known, though
we have to keep in mind that the differences in the observing
conditions and the way the threshold pulse intensity values are chosen or
defined could influence such comparisons.
One additional feature of our data is that the distribution of giant pulses is
not uniform. There is a large fluctuation in the giant pulse occurrence
rate from one day to other. Almost all ($\stackrel{>}{_{\sim}} 99 \%$) of
the total giant pulses seen by us occurred during 35 or so
``active'' days out of our 141 days of observations, i.e., about one fourth fraction. 
Even out of these there were about a dozen particularly active
days where more than five percent of the total pulses observed were giants.
On these particular days PSR B0950+08 exhibited the highest rate of occurrence of giant
pulses seen from any pulsar. On the other hand
$\sim 20 \%$ of the days (27 out of 141) showed only ``quasi--nulls'' when no
detectable emission was seen either in the folded pulse data or in the power spectrum. 
No giant pulse was generally seen on these silent days and the mean flux density of the 
pulse on the day (for 32 minutes of observations) was below $\sim$ 0.3 Jy, 
implying that generally the pulse intensities on such days are at least an 
order of magnitude below the normal average pulse intensity of 3 Jy. 
Since the fraction of giant pulses and quasi--nulls seems to vary so drastically
from one day of observation to the other, the conflicting nulling fraction 
reported by  Smith (1973) and Hesse and Wielebinski (1974) on one hand and
by Ritchings (1976) on the other, could simply be due to their different 
epochs of observations. 

Figure~1 shows a sample of four days of consecutive observations. Figure~2
shows the corresponding power spectra. We notice that the giant pulse
activity changes very drastically from one day of observing to the other.
In order to quantitatively compare the variation in giant pulse activity from
one day to the other, we made use of the fact that there is almost a
 one to one correspondence between the rate and strength of giant pulses seen on each day and
the daily average pulse intensity value. This is seen in Figure~3 which shows 
a plot of the daily average pulse intensity values obtained from the average folded pulse profiles, 
generated using the precomputed pulse period for our telescope location and time of the 
day's observations. From Figure~3 we see that the daily average between day numbers 54 and 59 
mimics the change in giant pulse activity between 23/02/98 and 28/02/98 seen in Figures~1 and 2.
The plot shows that starting from a quiet phase, when there is hardly any
detectable pulse emission, there is a build up time of 1-3 days for the pulsar to
change into an active phase, with large average pulse intensity, which
is followed by a similar time interval of ``decay'' to the quiet level.
And such behavior seems to repeat again and again. Whether there is any 
quasi-periodicity in this cycle, on say, a 3--4 days period, needs to be 
checked. But for that one has to first make sure of the time scales of  
variations which could actually be smaller
as our interpretation is based on the observations taken through a
half-an-hour time window every 24 hours of sidereal time.
Both the very large percentage of
giant pulses on some days and the frequent switching between
days of extreme quietness and of giant-pulse activity makes PSR B0950+08
perhaps the most violently variable among the known pulsars.

The process that gives rise to {\em individual} giant
pulses seems to be quite erratic. Even on days of extreme
giant pulse activity, when more than five percent of all pulses are giants, 
the giant pulses seemed to occur quite randomly. It is only one third of the time that a couple of giant 
pulses followed each other in quick succession. Most of the time, giant 
pulses appeared in isolation with pulses on either side being more than an order of magnitude lower in
intensity, and often falling below the noise level. 
There were large quiet intervals, often extending to many 
pulse periods, when no giant pulse was seen. Such behavior was seen to repeat on many days.

In both the Crab pulsar and PSR B1937+21, the cumulative distributions
of the giant pulse intensity, $S$, have been shown to follow a power law
$N(>S) \propto S^{\alpha}$ (Argyle \& Gower 1972; Lundgren et al. 1995; 
Bhat, Tingay \& Knight 2008; Cognard et al. 1996; Kinkhabwala \& Thorsett 2000).
For PSR B0950+08 we have examined the cumulative distribution of pulse
intensity on the days of giant pulse activity. As an example we show in
Figure 4 a plot of the normalized cumulative distribution of pulse
intensities for PSR B0950+08 for three different days. Also shown are the 
power-law fits to the observed distributions, which yield the best-fit values
for the index $\alpha$ to be in a tight range around $-2.2 \pm 0.2$.
This value for $\alpha$ is quite similar to those found in the case of the 
Crab pulsar ($\alpha \simeq -2.3$; Lundgren et al. 1995) and PSR B1937+21 
($\alpha \simeq -1.8$; Cognard et al. 1996).
In fact there may even be further resemblance between the giant pulse
distributions in PSR B0950+08 and the Crab pulsar. In figure 4 we notice that
the cumulative distribution at less than 30 times the average intensity
falls below the fitted line on all three days, perhaps indicating
a departure from the power-law. If we ignore the lower pulse intensity points 
and make a fit only to the higher intensity ($>30$ times the average) then the power 
law index is steeper, close to $-2.7 \pm 0.2$. This trend is quite similar to that seen
in the case of the Crab pulsar, where the slope tapered off at intensities  
close to the giant pulse threshold level (Lundgren et al. 1995; Karuppusamy et al. 2010). 
Thus the process responsible for the giant pulse phenomena might be rather similar in such cases.

Cairns et al. (2004) have found B0950+08 to emit giant micropulses which according to them 
do not meet the criteria of being called giant pulses, as their pulse-integrated flux does not exceed 10 times 
the average pulse-integrated flux. We may point out that as discussed above, the giant pulse activity in 
B0950+08 seems to vary so much from day to day that it will not be surprising if observations lasting for only 
about half an hour on a single day (7072 pulses of Cairns et al. 2004), which is in fact very close to the time 
interval for our daily observations, does not show very large pulses. Our observations which represent the 
pulse-integrated flux analysis do show the individual pulse intensity to exceed the average value by more 
than a factor of 10 in $\sim5\%$ of the total pulses on ``active days''.

\section{Discussion}
The origin of these giant pulses can be traced to the pulsar itself. One can
unambiguously rule out any instrumental effects like large gain fluctuations
of the receiver system masquerading as giant pulses. The calibrated
noise level in the records does not fluctuate by any appreciable amount from
one day to the other (Fig. 1) irrespective of the level of
the giant-pulse emission seen. Nor could these be any interference spikes
as the periodicity of the pulsar is so clearly visible in the spectral
plots in Figure~2. One can also rule out any ionospheric or
interplanetary scintillation effects as the cause since our
observations have been carried out both during day and night times,
depending upon the time of transit of the source during different months
over the year, and no systematic differences in the giant-pulse emission
rates are found in the records.

The fluctuations in the pulse intensities on different time scales
have also been explained by the diffractive and refractive
scintillations of the interstellar medium; especially, the
long term changes in flux density on time scales of days or longer have been explained
in terms of RISS (refractive interstellar scintillation;
Rickett, Coles \& Bourgois 1984). The time scales for the build up of
individual giant pulses, as observed by us, are too fast for the
scintillation effects to explain them. As we have already mentioned,
in the case of giant pulses, the pulse intensity is often larger than that of
the neighboring pulses by much more than an order of magnitude. These
enhancements in intensity sometimes last over a few consecutive
pulses, but more often than not these enhancements may have much shorter
durations ($\stackrel{<}{_{\sim}}$ a pulse period) and with the consecutive
giant pulses quite often
separated by many pulse periods. Thus the rise/fall time for the giant
pulse emission in our observations is at most up to a few pulse periods,
which is many orders of magnitude smaller than the time scales of
scintillation. The diffractive time scales for PSR B0950+08 at 103 MHz are
estimated to be about 10 minutes, while the time scale of 
RISS is expected to be around 47 days, calculated from the values
given for 74 MHz by Gupta , Rickett \& Coles (1993).
Thus these extreme intensity enhancements of individual giant pulses
cannot be explained by diffractive or refractive scintillation 
and it can therefore be construed that the giant pulses seen by us 
are intrinsic to the pulsar. 

Even though some presence of interstellar scintillation in our data 
cannot be ruled out, none of the variations that we have reported here can
be wholly explained by the scintillation.
We may point out that a slow intensity variation visible in Figure~1,  
peaking near the middle of each day's plot, is merely the 
east-west primary beam pattern of our telescope, which is a transit system,
and that corrections for this primary beam pattern were made before any further analysis.
The observed variability time scale of 3--4 days in Figure 3 is too slow for
diffractive scintillation but too fast for refractive. 
The change in the daily average by a factor of $\stackrel{>}{_{\sim}}70$
(from about $\stackrel{<}{_{\sim}}0.3$ to $\stackrel{>}{_{\sim}}20$ Jy)
just in a day's time,
does not fit with the expected refractive time scales. These day
to day variations in the average intensity, and the giant pulse rate, 
appear to be among the most extreme for pulsars. For example, in the
case of the Crab pulsar, Lundgren et al. (1995) have seen day to day intensity
variability to be $\approx 50 \%$ of the average
flux density, along with a factor of two change in the rate of giant pulses
observed, and these they have attributed to RISS, which has a
characteristic time scale of 2-5 days for their observing parameters.
On the other hand for PSR B0950+08 the variations in the daily--average 
intensity are about an order of magnitude as compared to the predicted modulation
index of $\approx 0.22$ (implying expected rms fluctuations of about $22\%$
in intensity) at 103 MHz (see Gupta et al. 1993). The fluctuations in the
giant pulse rate are equally drastic with some days showing hardly any
unambiguous giant pulses, while the very next day may show as many as
300-400 giant pulses during our daily half an hour of observing.
These extreme variations are much beyond the standard scintillation
predictions.

The use of a daily mean instead of the global mean for defining a threshold
level for giant pulses may appear preferable. However, a problem in the use of 
daily mean is that whenever there will be a large fraction of genuine giant 
pulses, the daily mean value will also accordingly go up (even in the absence of 
scintillation). In fact if we mask out the giant pulses (which are only a small 
fraction of the total number of pulses, at most 5 percent on very active days), 
then the daily mean drops substantially closer to the global mean. We may add here 
that even if we use a ``pulse--on'' daily mean, still we find a fairly large number of pulses in 
our data that exceed the mean by more than an order of magnitude, something
rarely ever seen in other pulsars. For example on 28/02/98 (Fig.~1 and 3) the daily mean is 
about 25 Jy, and there are about 20 pulses with a flux density higher than 250 Jy, 
that is ten times that day's ``pulse--on'' mean. Even on 23/02/98 the daily mean is about 7 Jy, 
and there are again about 50 pulses with flux density higher than 70 Jy on that day. Further, as 
we mentioned earlier, most of the giant pulses occur in isolation without 
another giant pulse in close proximity. Both these facts imply that it is not 
the overall level of the pulsar intensity that got boosted up by scintillation. 
In spite of it, if we were nonetheless to ascribe daily variations to some
{\em extreme} interstellar scintillation effects, and to then assume that
an increase in the number of giant pulses observed above a threshold on some
days is merely a result of an increase in the overall pulse intensity due to
scintillation, it will still be very hard to explain the presence of large
variations {\em within} this scintillation-enhanced intensity (by more than an order
of a magnitude) in at most a few pulse periods (less than a second of time!).
Such fluctuations have never been seen earlier in other pulsars (except of course 
in giant--pulse emitting pulsars), and
are too fast to be due to scintillation. The daily variations that result from 
the fluctuation in the contribution of the giant pulse intensity too are then intrinsically 
arising from the pulsar, with the interstellar scintillation playing only a minor role, if any.  

In order to make sure that in spite of the shortcomings,  
like the absence of full polarization, a large samplig rate of 48 ms 
and the pulse smearing ($\sim 36$ ms) across the band due to dispersion,  our Rajkot data are still meaningful, 
we observed this pulsar with the Westerbork Synthesis Radio Telescope (WSRT) using the 
Pulsar Machine (PuMa, Strom 2002) at 297 MHz. 
We also observed the pulsar with the Ooty Radio Telescope (ORT, Swarup et al. 1970) at 327 MHz 
for a number of sessions lasting a few hours each.
In both WSRT and ORT observations (Singal, Manoharan \& Strom 2002) a trend similar to as seen at 
Rajkot was noticed. It seems that both the frequency and the strength of the giant pulses 
may vary over a few hour time scale at these frequencies. There are stretches where 
pulses almost disappear (quasi--nulling!), to be followed within several hours by pulses as much as two orders 
of magnitude above the average pulse strength. These time scales appear to be too short 
for either the refractive or diffractive interstellar scintillation, and we see a similar 
pattern at frequencies which differ by a factor of three. 
Figure~(5a) shows a record from the WSRT while figure~(5b) shows one from the Rajkot Telescope (both records 
belong to different dates).  
A comparision of giant pulses in (5a) and (5b) shows them to be quite similar.  
The WSRT data (figure~(5a)) has full polarization, millisec time resolution, and the data have 
been dedispersed to remove any pulse smearing across the bandwidth. The similarity of the giant 
pulse activity in the two records gives us confidence that our Rajkot data are reasonably reliable.
Figure (6) shows a record of WSRT data where the giant pulses seem to suddenly get ``switched off''
within at most a few pulse periods, i.e., within about a second. 
The expected time scales of such variations from interstellar scintillations (ISS) are  
many days, more than 5 - 6 orders of magnitude larger than the actually observed 
time scales. Thus it is unlikely these ``switching off'' of giant pulses is due to ISS.

It is likely that giant pulses 
comprise one or more micro--pulses which may be intrinsically extremely bright (Hankins 1971). 
Popov et al. (2002) and Cairns et al. (2004) found microstructure with characteristic 
time scales in sub-millisecond range. 
At least some individual giant pulses observed in B0950+08 with the WSRT at 297 MHz remained unresolved 
at millisecond time scales (Figure~7),  
which points to an origin of this giant pulse activity to be an intrinsic phenomenon within the pulsar.

From the data analysis based on a couple of simultaneous observations at 103 MHz at Rajkot and 
at 297 MHz at Westerbork, there does not appear to be an obvious 
correlation between the occurrence of individual gaint pulses, or even of pulse strength 
when averaged over minute time scales. It may well be that the giant pulse activity is not 
very broadband, and indeed the WSRT PuMa data shows decorrelation of individual pulses 
within the 10 MHz band. This might agree with a similar conclusion for B1937+21 that the frequency 
bandwidth of the individual giant radio pulses is relatively narrow (Popov \& Stappers 2003).
Further observations of these giant pulses with a 
better sensitivity and time resolution, and preferably at two or more 
frequency bands simultaneously, are needed to help us in discerning the nature and the ultimate 
origin of these giant pulses.

The nature of the giant pulses still remains obscure though there are some efforts in this 
direction (Cairns 2004; Petrova 2004). Equally puzzling is why this phenomenon 
takes place only in a handful of pulsars. 
Cognard et al (1996) examined the period $P$ and the period derivative $\dot{P}$ of a 
few dozen strongest pulsars and estimated the strength of their magnetic  field at the light
cylinder, $B_{lc} \sim 3 \times 10^{8}  P^{-2.5} \dot{P}^{0.5}$ G.
They noted a rather high value  $\sim 10^{6}$ G for the Crab pulsar
($P=33$ ms, $\dot{P}=10^{-12.4}$ sec/sec), as well as PSR B1937+21
($P=1.56$ ms, $\dot{P}=10^{-19}$ sec/sec), 
which compared to  other pulsars is an order of magnitude or more higher.
From this they suggested that the giant pulse phenomenon may have something to do with the strength
of the magnetic field at the light cylinder radius.  Even B1821-24 (Romani \& Johnston 2001) has 
a relatively high $B_{lc}\sim 7 \times 10^{5}$~G. On the other hand Vela (B0833-45), a giant micropulse emitter,  
does not have so high a $B_{lc}$ value ($\sim 4 \times 10^{4}$ G). From the
known parameters of PSR 0950+08, ($P=253$ ms, $\dot{P}=10^{-15.6}$ sec/sec), 
we find $B_{lc}$ to be only $\sim$ 150 G, and it does not seem to support the hypothesis  
that the giant pulse emission physics is particularly dependent on the high magnetic
field strength at the light cylinder.
It also shows that giant pulses are not all associated with fast pulsars alone.

The extremely high equivalent brightness 
temperatures of the giant pulses indicate that they originate from nonthermal, coherent 
emission processes (Hankins et al. 2003). 
A joint study at radio and $\gamma$-rays of the giant pulses from the
Crab pulsar (Lundgren et al. 1995) showed that the $\gamma$-ray  emission 
remains unchanged during giant radio pulse emission and hence, they
concluded that radio coherence is the primary, if not the sole,
mechanism for producing fluctuations in the radio emission.
They also concluded that the largest giant pulses are a sum of a
large ensemble of particularly dense coherently emitting regions
and the smaller giant pulses are formed by a smaller number of
less dense emission regions. According to this suggestion, it is the fluctuations 
in the number of charges partaking in the coherent radiation process that gives rise to 
the intense variation in the net radio emission of the pulse intensity. 
One can argue that the giant pulse emission and the nulling may be  
opposite manifestations of the same physical process in a pulsar (Singal 2001), while 
an enhanced degree of bunching of radiating charges may give rise to a high degree of coherence, 
resulting in giant pulses, a drop in the degree of coherence may cause nulling. 
\section{Conclusions}
Giant pulses were shown to arise from  B0950+08, a normal-period pulsar. 
The observation made at 103 MHz showed that the energy in these giant pulses 
exceeded that of the average pulse energy by much more than an order of magnitude. The 
cumulative intensity distribution of these giant pulses appears to follow a power law, with 
power-index $-2.2\pm 0.2$, very similar to that seen in other cases of giant pulses detected only in a few fast pulsars, 
indicating that nature of the giant pulses may be similar in the fast and normal period pulsars. 
\section*{Acknowledgements}
The WSRT observations were done by one of us (AKS) in collaboration with Richard Strom, whom we are grateful for 
his help in the WSRT data reduction. The WSRT is operated by ASTRON. Apoorva Singal's help is appreciated in the 
production of figures. We thank Sandeep Doshi for his valueable help in observations with the Rajkot telescope.
We also thank an anonymous referee for going through our manuscript diligently and suggesting changes for 
improvements in its readability.

\begin{figure}
\includegraphics[width=16.4cm]{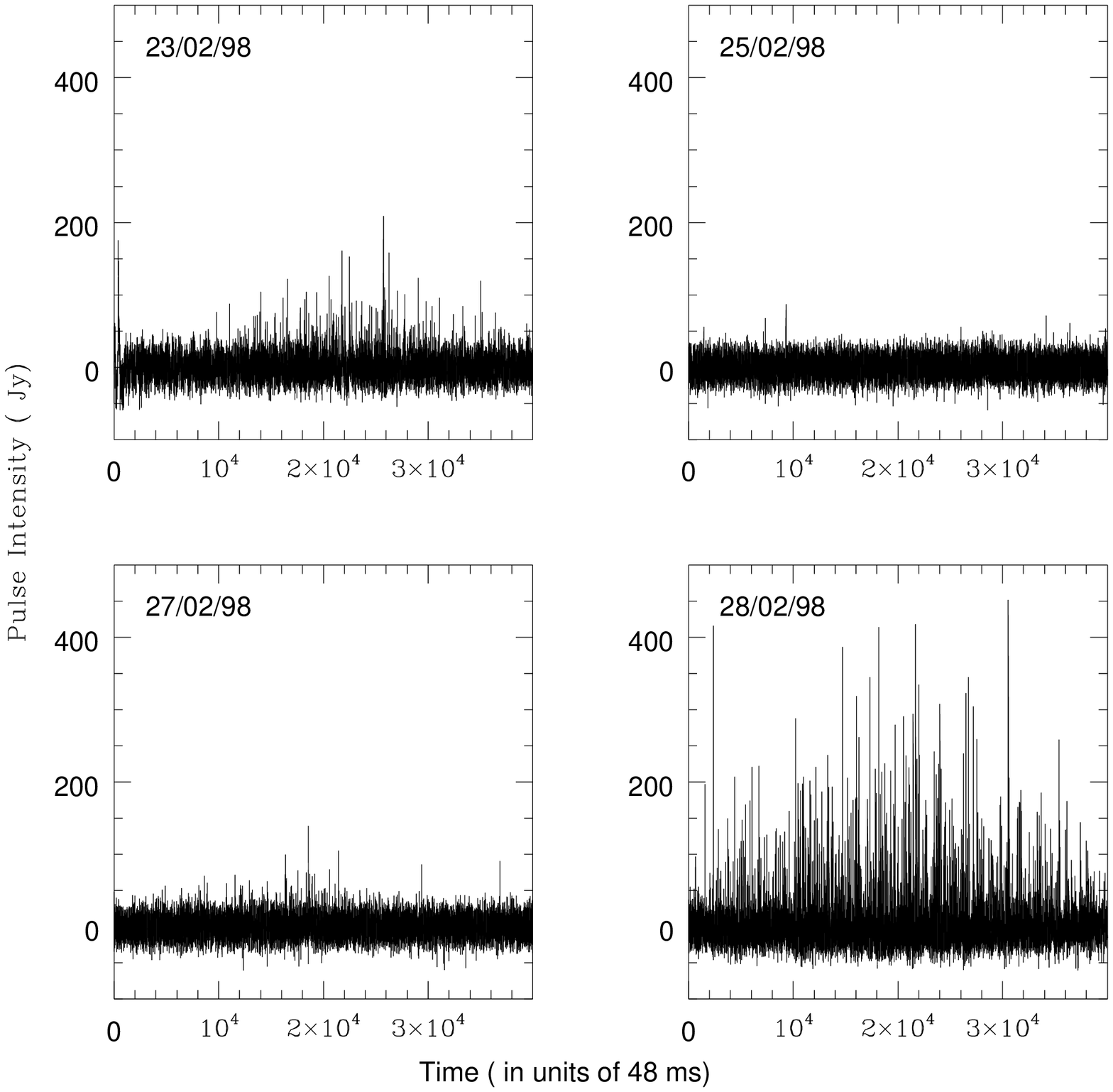}
\caption{A plot of four consecutive observations of PSR B0950+08, showing a total of
40,000 data points in each case. With sampling interval of 48 ms, each observation amounts to 
an observation time of 32 minutes corresponding to about 7600 pulses. A running average 
of 1000 data points has been subtracted throughout to remove any slow baseline drifts.}
\end{figure}

\begin{figure}
\includegraphics[width=16.4cm]{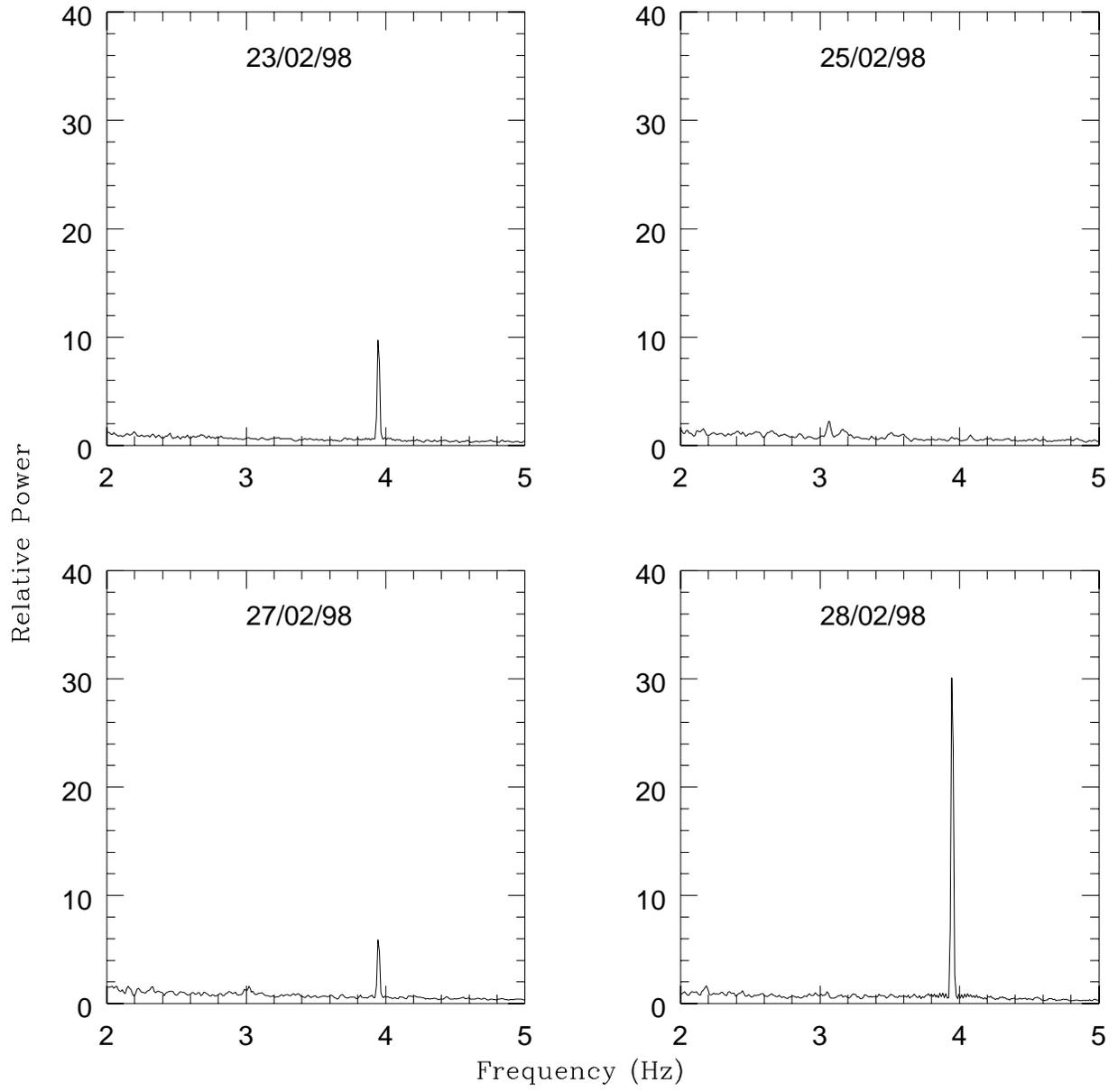}
\caption{Power-spectrum plots of the four observations of figure~1.}
\end{figure}

\begin{figure}
\includegraphics[width=16.4cm]{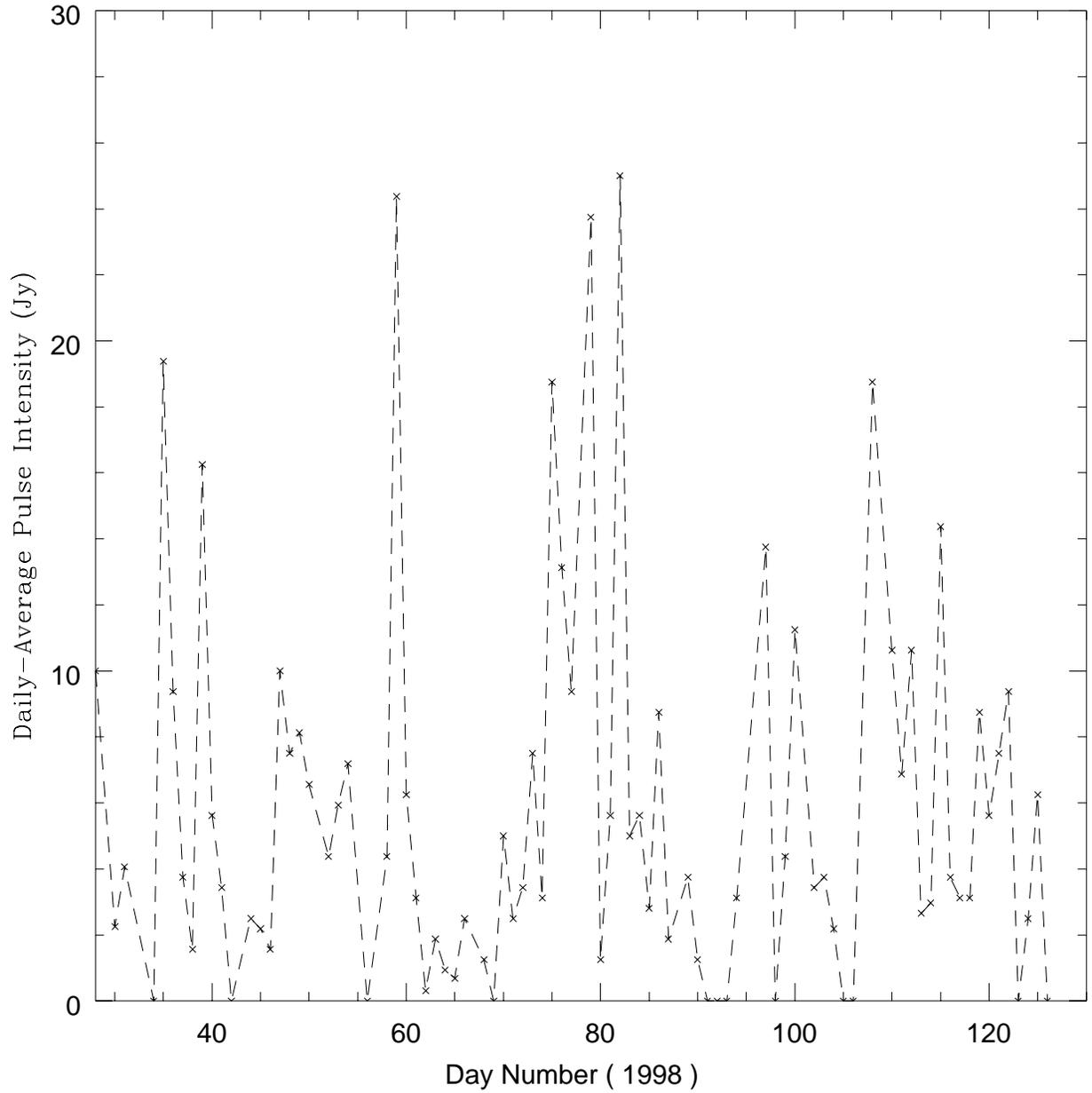}
\caption{A sample plot of the daily average pulse
intensities for 84 days of observations in the first half of 1998,
obtained mostly between February and April.}
\end{figure}

\begin{figure}
\includegraphics[width=16.4cm]{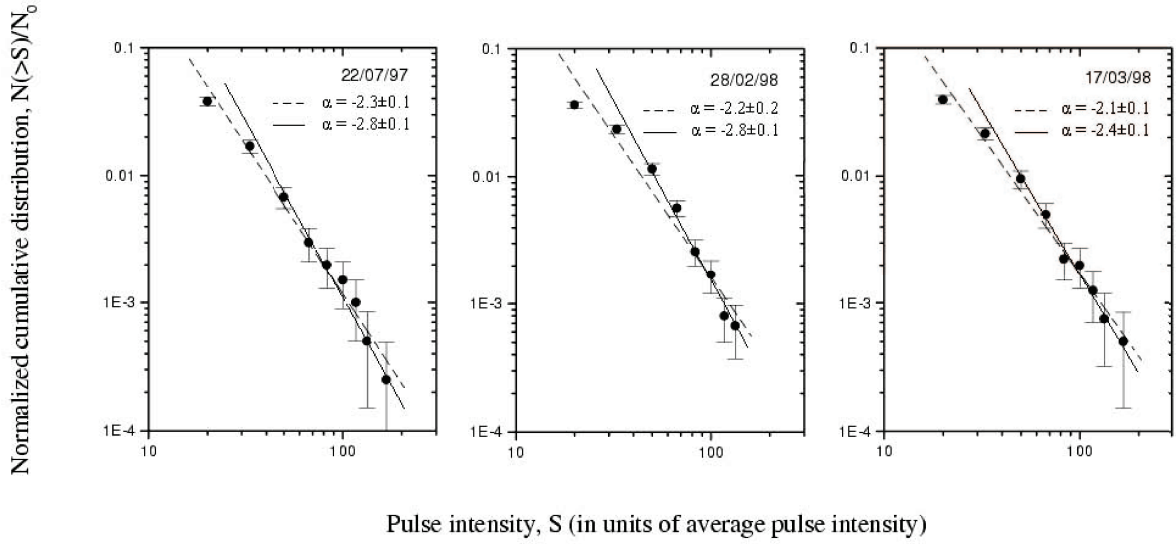}
\caption{Normalized cumulative distribution of the giant
pulse intensity shown in log-log plot for three different days of
observations. The dashed line in each case shows a
power-law fit to the overall observed distribution 
while the solid line shows a power-law fit to the distribution only at high intensity 
($>30$ times the average), with the derived slope $\alpha$ steeper in the latter case.}
\end{figure}
\begin{figure}
\includegraphics[width=16.4cm]{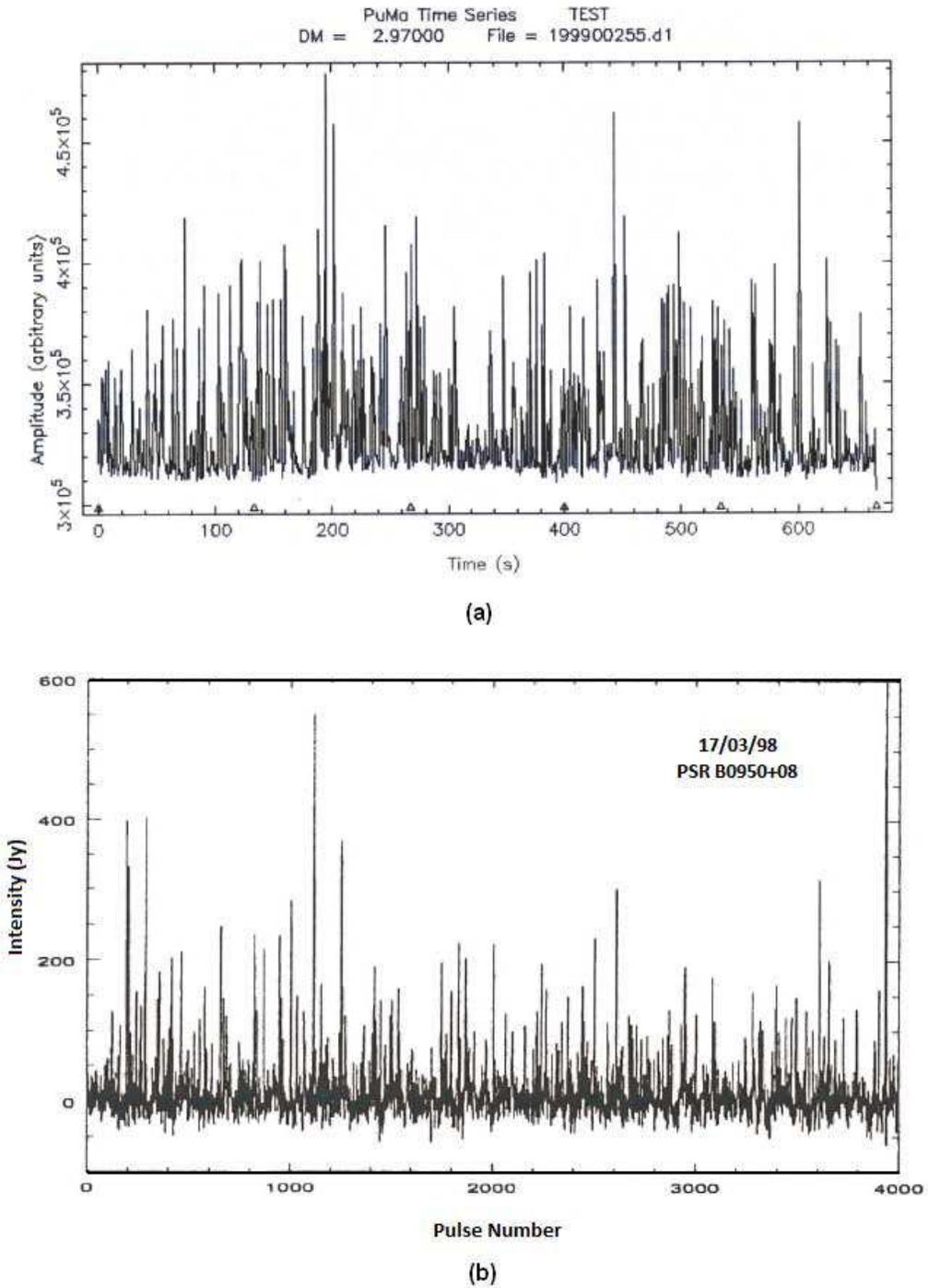}
\caption{(a) A WSRT record of giant pulses at 297 MHz on 12/09/99. (b) Data from the Rajkot Telescope on 17/03/98.  
The giant pulse distributions in the plots (a) and (b) appear to be quite similar.  
}
\end{figure}
\begin{figure}
\includegraphics[width=16.4cm]{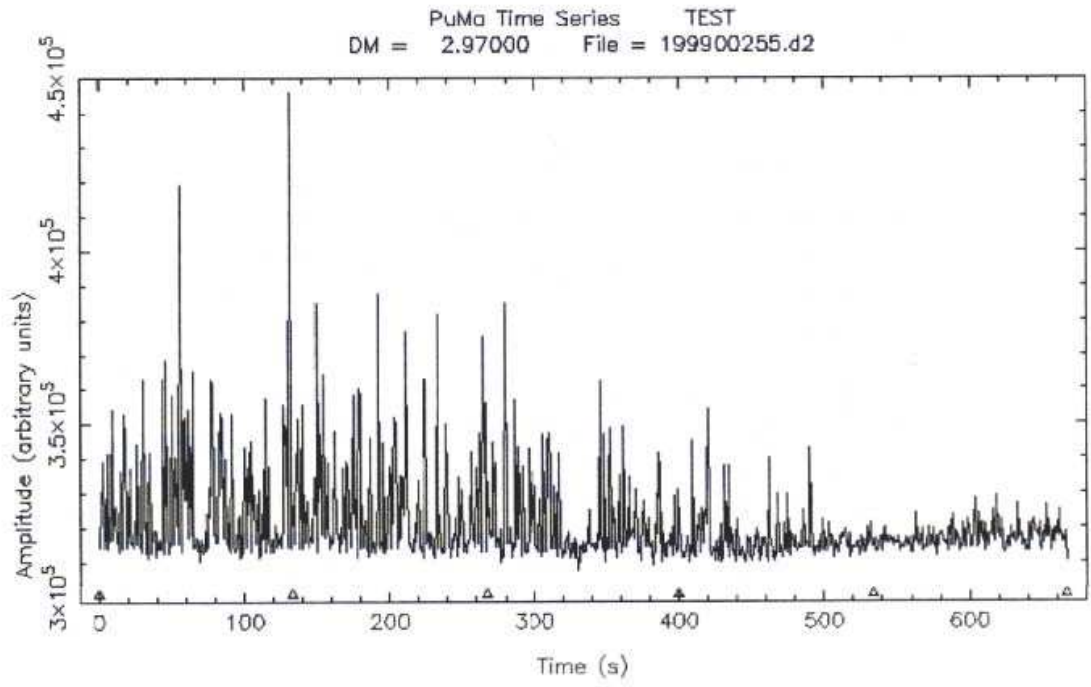}
\caption{A record of WSRT data where the giant pulses suddenly 
seem to get switched off.}
\end{figure}
\begin{figure}
\includegraphics[width=16.4cm]{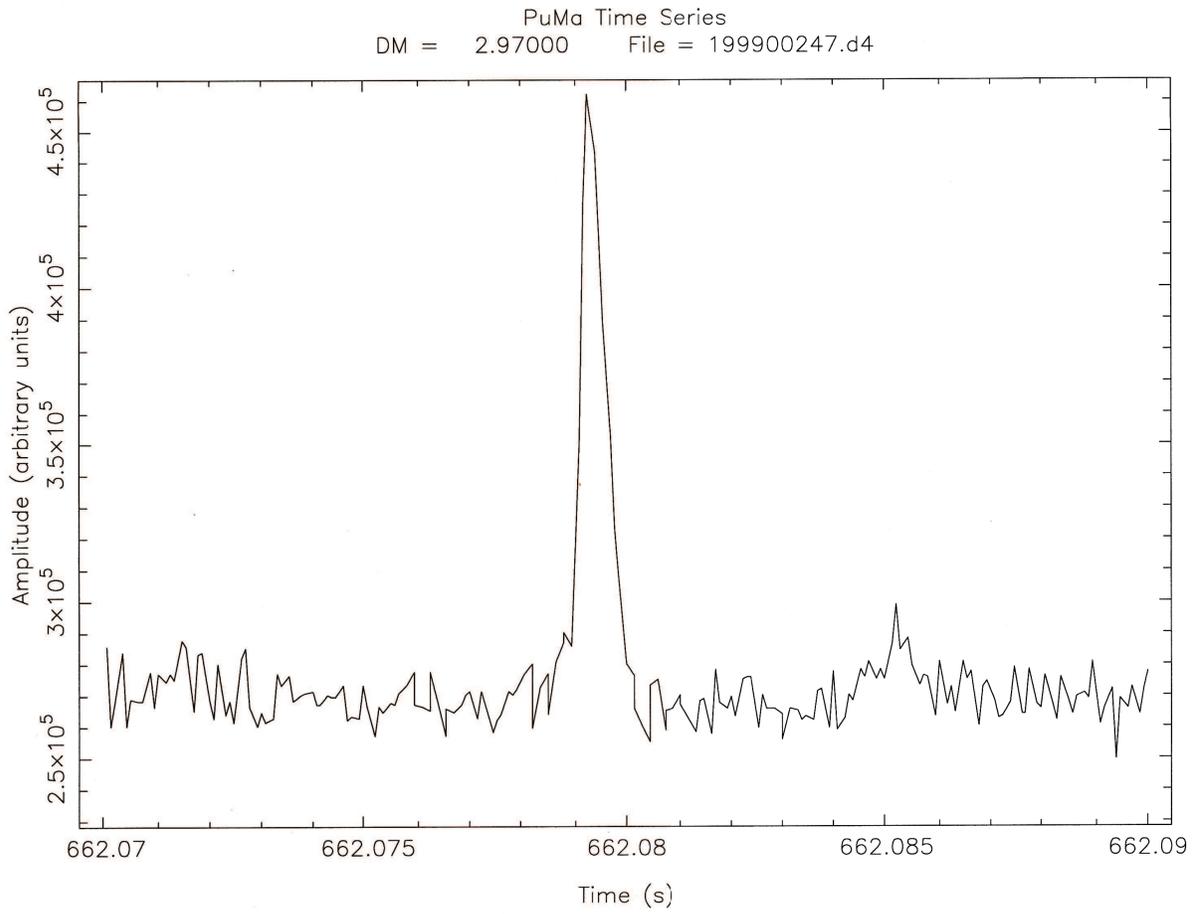}
\caption{A high time resolution of one of the giant pulses observed with the WSRT 
shows it to be unresolved at millisecond time scales.}
\end{figure}
\end{document}